\documentclass[aps,prb,showpacs,groupedaddress,twocolumn]{revtex4}

\usepackage{graphicx}
\begin{document}

\title{CT invariant quantum spin Hall effect in a ferromagnetic graphene
}

\author{Qing-feng Sun$^{1,\star}$ and X.C. Xie$^{1,2}$}
\affiliation{ $^1$Institute of Physics, Chinese
Academy of Sciences, Beijing 100190, China\\
$^2$Department of Physics, Oklahoma State University, Stillwater,
Oklahoma 74078 }

\date{\today}

\begin{abstract}
We predict a quantum spin Hall effect (QSHE) in the ferromagnetic
graphene under a magnetic field. Unlike the previous QSHE, this QSHE
appears in the absence of any spin-orbit interaction, thus, arrived
from a different physical origin. The previous QSHE is protected by
the time-reversal (T) invariance. This new QSHE is protected by the
CT invariance, where C is the charge conjugation operation. Due to
this QSHE, the longitudinal resistance exhibits quantum plateaus.
The plateau values are at $1/2$, $1/6$, $3/28$, ... , (in the unit
of $h/e^2$), depending on the filling factors of the spin-up and
spin-down carriers. The spin Hall resistance is also investigated
and is found to be robust against the disorder.
\end{abstract}


\maketitle

In the years since the spin Hall effect (SHE) has been discovered,
it has generated great interest.\cite{ref1,ref2,ref3,ref4} In the
SHE, an applied longitudinal charge current or voltage bias induces
a transverse spin current due to the spin-dependent
scatterings\cite{ref1,ref2} of the spin-orbit interaction
(SOI)\cite{ref3}. Soon afterwards, the quantum SHE (QSHE) is also
predicted.\cite{ref5,ref6} The QSHE occurs in a topological
insulator in which the bulk material is an insulator with two
helical edge states carrying the current.\cite{ref7} The edge
states, with opposite spins on a given edge or opposite edges for a
given spin direction containing opposite propagation directions,
lead to a quantized spin Hall conductance. The QSHE is a new quantum
state of matter with a non-trivial topological property. The
existence of QSHE was first proposed in a graphene film in which the
SOI opened a band gap and established the edge
states.\cite{ref5,ref6} But the sequent work found that the SOI in
the graphene was quite weak and the gap-opening was small, so the
QSHE was difficult to observe.\cite{ref8} Soon afterwards, the QSHE
was also predicted to exist in some other
systems.\cite{ref9,ref10,ref11,ref12} Recently, the QSHE was
successfully realized in the CdTe/HgTe/CdTe quantum wells, and a
quantized longitudinal resistance plateau was experimentally
observed due to the QSHE.\cite{ref11}

Another subject that has also been extensively investigated in
recent years is the graphene, a single-layer hexagonal lattice of
carbon atoms\cite{ref13} after it has been successfully
fabricated.\cite{ref14,ref15} The graphene has a unique band
structure with a linear dispersion near the Fermi surface, giving it
many peculiar properties. For example, the quasi-particles obey the
Dirac-like equation and have relativistic-like behaviors, and its
Hall plateaus are at the half-integer values.

In this Letter, we predict a new kind of QSHE in a ferromagnetic
graphene. Let us first imagine a two-dimensional system consisting
of the following characteristics: (i) its carriers contain electrons
and holes; (ii) both electrons and holes are completely
spin-polarized with opposite spin polarizations. When a high
perpendicular magnetic field is applied to the system, the edge
states are formed and the carriers move only along the edges. In
particular, the electrons (with their spins up) and holes (with
their spins down) move in opposite directions on a given edge (see
the inset in Fig.1a). Therefore, the QSHE automatically exists in
this system. Although the ordinary metals (or doped semiconductors)
cannot meet the above two characteristics, a ferromagnetic graphene
does. Recently, several approaches to realize a ferromagnetic
graphene have been suggested.\cite{ref16,ref17,ref18} For example,
the ferromagnetic graphene can be realized by growing the graphene
on a ferromagnetic insulator (e.g. EuO).\cite{ref17} For a
ferromagnetic graphene, as soon as the Fermi energy $E_F$ is tuned
to lie between the spin-up and spin-down Dirac points (see the inset
in Fig.1b), the above two characteristics are met and the QSHE
occurs. In the following calculations, we consider four- and
six-terminal graphene Hall bars (see the insets in Fig.1a). The
results reveal that the transverse spin current and spin Hall
resistance indeed show the quantized plateaus because of the QSHE.

Comparing this new QSHE with the previously-studied QSHE, there are
two essential differences: (i) The previous QSHE comes from the SOI
and the proposed systems all contain the time-reversal
symmetry,\cite{ref5,ref6,ref7,ref9,ref10} while the present QSHE
exists without the SOI and breaks the time-reversal symmetry.
However, this new QSHE is protected by CT invariance. (ii) In the
previous QSHE, the edge states only carry a spin current while at
equilibrium; in this QSHE system, the edge states carry both spin
and charge currents at equilibrium with the two edges states being
CT partners of each other (see the inset in Fig.1a). Thus, this is a
new kind of QSHE and the system is a new type of topological
insulator. Due to the topological invariance, the plateaus of the
spin Hall resistance are robust to disorder or impurity scattering.
So the plateau is very stable and its value can be used as the
standard value for the spin Hall resistance.

\begin{figure}
\includegraphics[width=8.5cm,totalheight=7.5cm]{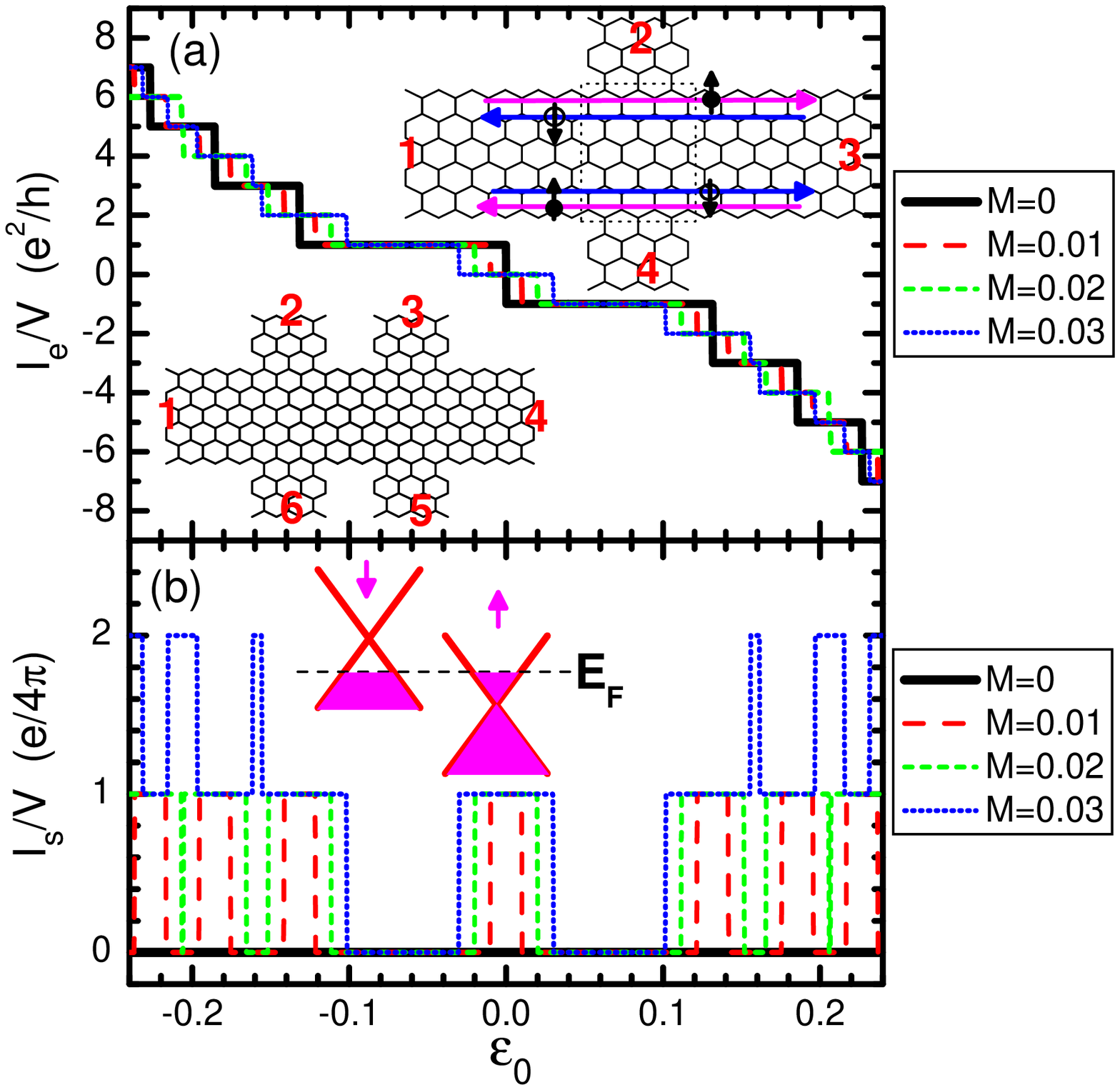}
\caption{ (Color online) The Hall conductance $I_e/V$ (a) and spin
Hall conductance $I_s/V$ (b) vs. the energy $\epsilon_0$ for $N=80$
and $\phi=0.005$. The two insets in (a) are the schematic diagram
for the four- and six-terminal graphene's Hall bars. The inset in
(b) is the schematic diagram for band structure of the ferromagnetic
graphene while $\epsilon_0 +M > E_F>\epsilon_0 -M$. }
\end{figure}

In the tight-binding representation, the four- or six-terminal
ferromagnetic graphene device (see the insets in Fig.1a) can be
described by the Hamiltonian:\cite{ref19}
\begin{equation}
  H = \sum_{i,\sigma} (\epsilon_0-\sigma M) a^{\dagger}_{i\sigma}
  a_{i\sigma}
      -\sum_{<ij>, \sigma} t e^{i\phi_{ij}} a_{i\sigma}^{\dagger} a_{j\sigma},
\end{equation}
where $a_{i\sigma}$ and $a_{i\sigma}^{\dagger}$ are the annihilation
and creation operators at the discrete site $i$. $\epsilon_0$ is the
on-site energy (i.e. the Dirac-point energy), $M$ is the
ferromagnetic exchange split\cite{ref17}, and $t$ is the nearest
neighbor hopping element. Here, the whole device, including the
center region and four or six terminals, is made of the
ferromagnetic graphene. With the presence of a perpendicular
magnetic field $B$, a phase factor $\phi_{ij}$ is added to the
hopping element, $\phi_{ij}=\int_i^j \vec{A} \cdot d\vec{l}/\phi_0$
with the vector potential $\vec{A}=(-By,0,0)$ and $\phi_0=\hbar/e$.

The transmission coefficient $T_{pq\sigma}(\epsilon)$ from the
terminal $q$ to the terminal $p$ with spin $\sigma$ can be
calculated from the equation:\cite{ref20} $T_{pq\sigma}(\epsilon) =
Tr [{\bf \Gamma}_{p\sigma} {\bf G}_{\sigma}^r {\bf \Gamma}_{q\sigma}
{\bf G}^a_{\sigma} ]$, where ${\bf \Gamma}_{p\sigma}(\epsilon)=
i[{\bf\Sigma}^r_{p\sigma}(\epsilon)
-{\bf\Sigma}^a_{p\sigma}(\epsilon)]$, the Green functions ${\bf
G}^r_{\sigma}(\epsilon)=[{\bf G}^{a}_{\sigma}(\epsilon)]^{\dagger}
=1/[\epsilon-{\bf
H}^{cen}_{\sigma}-\sum_{p}{\bf\Sigma}^r_{p\sigma}]$, and ${\bf
H}^{cen}_{\sigma}$ is the Hamiltonian of the center region. The
retarded self-energy ${\bf\Sigma}^r_{p\sigma}(\epsilon)$ due to the
coupling to the terminal $p$ can be calculated
numerically.\cite{ref21} After obtaining the transmission
coefficient, the particle current in the terminal $p$ with the spin
${\sigma}$ can be calculated from the Landauer-B$\ddot{u}$ttiker
formula: $I_{p\sigma}= (1/h)\int d \epsilon \sum_q
T_{pq\sigma}(\epsilon)[f_{q\sigma}(\epsilon)-f_{p\sigma}(\epsilon)]$,
where $f_{p\sigma}(\epsilon)
=1/\{\exp[(\epsilon-\mu_{p\sigma})/k_BT]+1\}$ is the Fermi
distribution function in the terminal $p$, with the spin-dependent
chemical potential $\mu_{p\sigma}$ and the temperature $T$. In the
following numerical calculations, we take $t=1$ as the energy unit
and only consider the zero temperature case ($T=0$), as the thermal
energy $k_BT$ is normally much smaller than other energy scales in
the problem. The sample width is denoted by $N$, and the insets of
Fig.1a show a system with $N=3$. In the calculations, we choose
$N=80$ and $40$, and the corresponding widths are $33.9nm$ and
$16.9nm$. The magnetic field is described by the $\phi$ with $\phi
\equiv (3\sqrt{3}/4) a^2 B/\phi_0$ and the magnetic flux in a
honeycomb lattice is $2\phi$.

We first consider the four-terminal device (see the inset at the top
right corner of Fig.1a) and a small bias $V$ is applied between the
longitudinal terminals 1 and 3 to study the induced charge current
$I_{ne}$ [$I_{ne}\equiv e(I_{n\uparrow}+I_{n\downarrow})$] and spin
current $I_{ns}$ [$I_{ns}\equiv
(\hbar/2)(I_{n\uparrow}-I_{n\downarrow})$] in the transversal
terminals 2 and 4. Here the boundary conditions for the four
terminals are $\mu_{1\uparrow}=\mu_{1\downarrow}=eV/2$,
$\mu_{2\uparrow}=\mu_{2\downarrow}=0$,
$\mu_{3\uparrow}=\mu_{3\downarrow}=-eV/2$, and
$\mu_{4\uparrow}=\mu_{4\downarrow}=0$. The currents in the terminals
2 and 4 satisfy the relations: $I_{2e}=-I_{4e}\equiv I_e$ and
$I_{2s}=-I_{4s}\equiv I_s$. Fig.1a and 1b show the Hall conductance
$I_e/V$ and spin Hall conductance $I_s/V$ versus the Dirac-point
energy $\epsilon_0$, respectively. For a non-ferromagnetic graphene
($M=0$) under the high magnetic field ($\phi=0.005$), $I_s/V$ is
zero and $I_e/V$ exhibits the plateaus at odd integer values
$ne^2/h$ ($n=\pm 1$, $\pm 3$, ...) due to the quantum Hall effect
(QHE). These results have been observed in recent
experiments.\cite{ref14,ref15} For a ferromagnetic graphene with
$M\not=0$, the spin current emerges (see Fig.1b) since the QSHE. The
spin Hall conductance $I_s/V$ also shows the quantized plateaus. By
considering the edge state under the high magnetic field, the
plateau values of $I_s/V$ and $I_e/V$ can be analytically derived to
be at $(\nu_{\uparrow}-\nu_{\downarrow})e/8\pi$ and
$(\nu_{\uparrow}+\nu_{\downarrow})e^2/2h$,\cite{supplement} where
$\nu_{\sigma}$ is the Landau filling factor for spin $\sigma$. In
particular, when $|\epsilon_0|<|M|$, in which case the Fermi energy
$E_F$ ($E_F=0$) is located between the spin-up Dirac-point
$\epsilon_0-M$ and the spin-down Dirac-point $\epsilon_0+M$, $I_e$
is zero and a net quantum spin current emerges in the transversal
terminals. In addition, if in the open circuit case, the spin
accumulation emerges at the sample boundaries instead of the spin
current.\cite{supplement}

Since the QSHE can give rise to quantum plateaus in resistances, we
next study the longitudinal and Hall resistances in the six-terminal
Hall device (see the inset in the lower left corner of Fig.1a). Now
we consider a small bias $V$ applied to the longitudinal terminals 1
and 4. The transversal terminals 2, 3, 5, and 6 are all voltage
probes, their charge currents vanish ($I_{pe}=0$) and
$\mu_{p\uparrow}=\mu_{p\downarrow}\equiv \mu_p$. Combining these
boundary conditions with the Landauer-B$\ddot{u}$ttiker formula, the
voltages $V_p$ ($V_p = \mu_p/e$) in four voltage probes can be
obtained, then the longitudinal resistance
$R_{14,23}=(V_2-V_3)/I_{14}$ and Hall resistance
$R_{14,26}=(V_2-V_6)/I_{14}$ are calculated, here $I_{14}= -I_{1e}=
I_{4e}$. The resistances contain the properties
$R_{14,26}=R_{14,35}$ and $R_{14,23}=R_{14,65}$.

\begin{figure}
\includegraphics[width=7cm,totalheight=8.5cm]{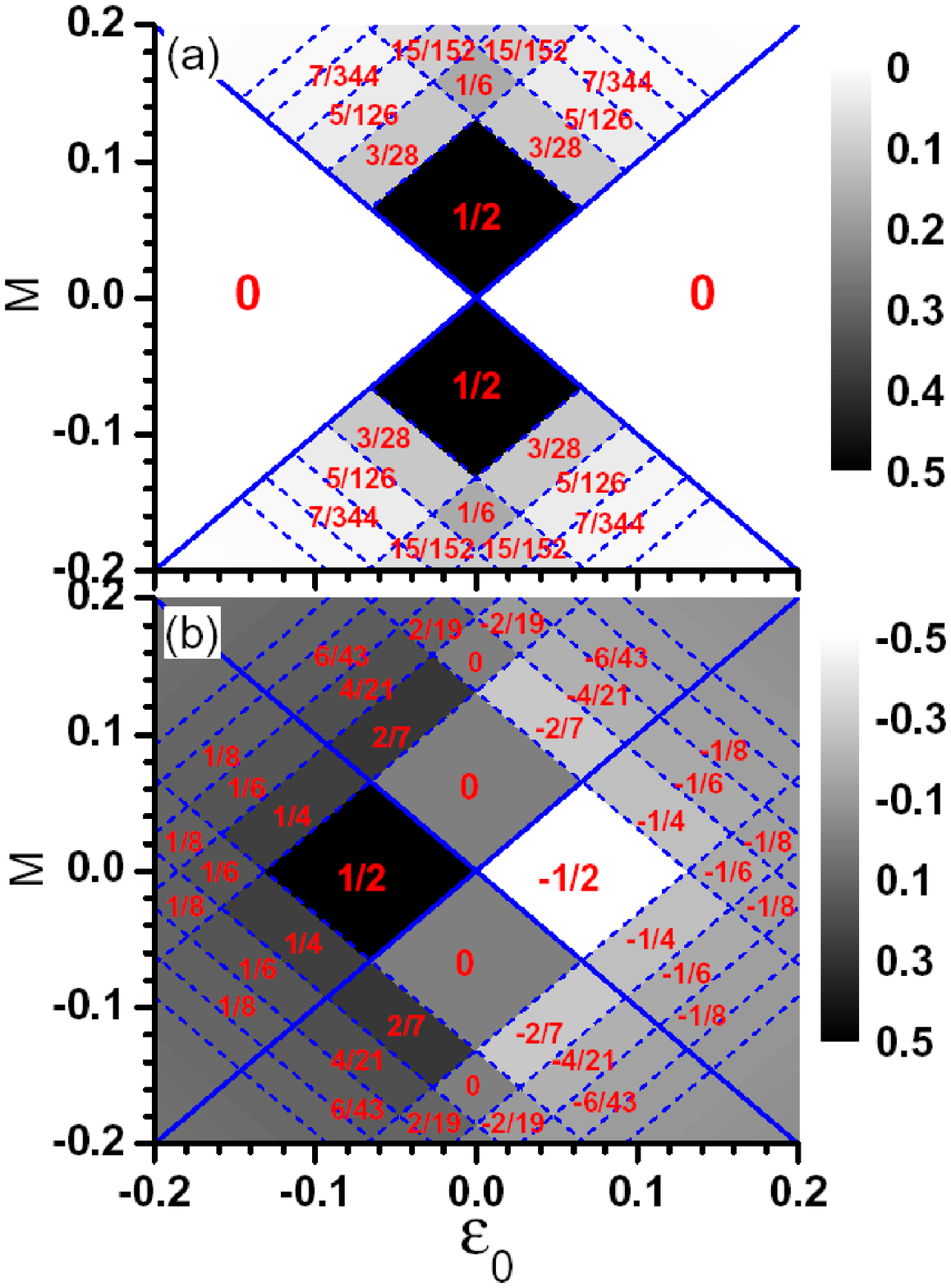}
\caption{ The panels (a) and (b) show the resistances $R_{14,23}$
and $R_{14,26}$ (in the unit of $h/e^2$) vs the exchange split $M$
and energy $\epsilon_0$ for $N=80$ and $\phi=0.005$. } \label{fig:2}
\end{figure}

Fig.2a and 2b show the longitudinal and Hall resistances,
$R_{14,23}$ and $R_{14,26}$, versus the energy $\epsilon_0$ and the
exchange split $M$ at an external magnetic field $\phi=0.005$. Due
to the QSHE and QHE, both $R_{14,23}$ and $R_{14,26}$ may be
non-zero, and they both exhibit plateau structures. The plateau
values are determined by the filling factors $\nu_{\uparrow}$ and
$\nu_{\downarrow}$. For the fixed filling factors $\nu_{\uparrow}$
and $\nu_{\downarrow}$, $R_{14,23}$ and $R_{14,26}$ maintain their
plateau values regardless of $\epsilon_0$ and $M$. By considering
the carriers transport along the edge states, the plateau values can
be analytically derived:\cite{supplement} $R_{14,23}=0$ and
$R_{14,26} = [1/(\nu_{\uparrow}+ \nu_{\downarrow})] h/e^2$ for
$(\nu_{\uparrow},\nu_{\downarrow})=(+,+)$ or $(-,-)$, and
$R_{14,23}= [|\nu_{\uparrow}\nu_{\downarrow}|/ ( |\nu_{\uparrow}|^3
+|\nu_{\downarrow}|^3 )] h/e^2$ and $R_{14,26} =
sign(\nu_{\uparrow})[ (|\nu_{\uparrow}|^2- |\nu_{\downarrow}|^2)/ (
|\nu_{\uparrow}|^3 +|\nu_{\downarrow}|^3 )] h/e^2$ for
$(\nu_{\uparrow},\nu_{\downarrow})=(+,-)$ or $(-,+)$. Some plateau
values for low $\nu_{\uparrow},\nu_{\downarrow}$ have been labeled
in Fig.2. The numerical results in Fig.2 are in excellent agreement
with the analytic plateau values (the differences between them are
less than $10^{-6}$). Furthermore, $R_{14,23}$ and $R_{14,26}$ have
the following properties: While $|\epsilon_0|>|M|$ with
$(\nu_{\uparrow}, \nu_{\downarrow})= (+,+)$ or $(-,-)$, the
longitudinal resistance $R_{14,23}$ is zero and only the Hall
resistance $R_{14,26}$ exists because the spin-up and spin-down
carriers are simultaneously either electron-like or hole-like and
move in the same direction. On the other hand, while
$|\epsilon_0|<|M|$ with $(\nu_{\uparrow}, \nu_{\downarrow})= (+,-)$
or $(-,+)$, the Fermi energy $E_F$ is located between $\epsilon_0+M$
and $\epsilon_0-M$, the longitudinal resistance $R_{14,23}$ emerges
since now the spin-up and spin-down carriers move in opposite
directions for a given edge. (i) While
$\nu_{\uparrow}=-\nu_{\downarrow} \equiv \nu$, the Hall resistance
$R_{14,26}=0$, only the longitudinal resistance $R_{14,23}$ exists
with the value $(1/2\nu)h/e^2$. This means that only the QSHE
emerges and the QHE vanishes in this region. In this case, the
system has the CT invariance. Furthermore, if
$\nu_{\uparrow}=-\nu_{\downarrow}=\pm 1$, $R_{14,26}=0$ and
$R_{14,23}=(1/2) (h/e^2)$. Now the observed phenomena are completely
the same with the QSHE from the SOI,\cite{ref5,ref6,ref7,ref9,ref10}
but their physical mechanisms are different. (ii) While
$\nu_{\uparrow}\not=-\nu_{\downarrow}$ but still with
($\nu_{\uparrow}$, $\nu_{\downarrow})=(+,-)$ or $(-,+)$, $R_{14,26}$
is now non-zero since the numbers of the spin-up and spin-down edge
states are different. In this case, both resistances $R_{14,26}$ and
$R_{14,23}$ have non-zero quantized plateaus and the QSHE and QHE
coexist. Fig.3 shows the resistances $R_{14,23}$ and $R_{14,26}$
versus the energy $\epsilon_0$ for a fixed $M$ (i.e. along the
horizontal lines in Fig.2), and it clearly shows that the quantum
plateaus persist very well.

\begin{figure}
\includegraphics[width=8.5cm,totalheight=4cm]{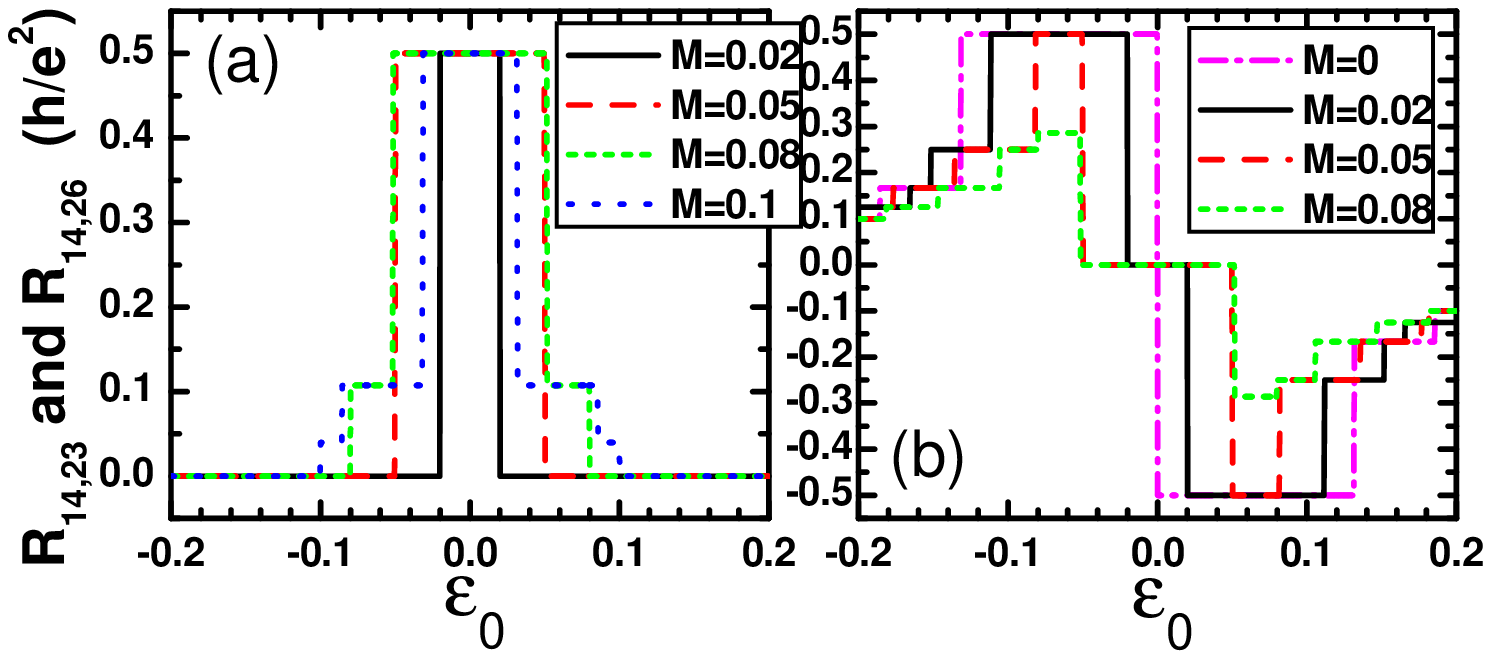}
\caption{ (Color online) The resistances $R_{14,23}$ (a) and
$R_{14,26}$ (b) vs the energy $\epsilon_0$ for $N=80$ and
$\phi=0.005$. } \label{fig:3}
\end{figure}

Up to now, we demonstrate the existence of QSHE in the ferromagnetic
graphene from both physical picture and detailed numerical
calculations. In the following, we study the properties of the spin
Hall resistance $R_s$, a measurable quantity robust to
dephasing\cite{ref22} and well reflecting the topological invariance
of the system. We again consider the four-terminal Hall bar. But now
the transversal terminals 2 and 4 are spin-biased probes with
boundary conditions $I_{p\uparrow} =I_{p\downarrow} = 0$ ($p=2,4$).
Here the spin Hall resistance $R_s$ is defined as the transversal
spin bias over the longitudinal charge current: $R_s \equiv
(\mu_{2\uparrow}-\mu_{2\downarrow})/eI_{13} =
-(\mu_{4\uparrow}-\mu_{4\downarrow})/eI_{13}$. Since the spin bias
$\mu_{n\uparrow}-\mu_{n\downarrow}$ is experimentally measurable, so
is the $R_s$.\cite{ref23,ref24} Fig.4 shows $R_s$ versus the energy
$\epsilon_0$ for different ferromagnetic exchange split $M$ and
magnetic field $\phi$. For $|\epsilon_0|>|M|$ with
$(\nu_{\uparrow},\nu_{\downarrow})=(+,+)$ or $(-,-)$, $R_s=0$. On
the other hand, while $|\epsilon_0|<|M|$ with
$(\nu_{\uparrow},\nu_{\downarrow})=(+,-)$ or $(-,+)$, $R_s$ exists.
$R_s$ exhibits the quantum plateaus, and its plateau values are at
$[1/(|\nu_{\uparrow}|+|\nu_{\downarrow}|)] h/e^2$. For a small $M$
(e.g. $M=0.02t$ or $0.05t$ in Fig.4b) or a high magnetic field
$\phi$ (e.g. $\phi=0.005$ in Fig.4a),
$(\nu_{\uparrow},\nu_{\downarrow})$ can only equal to $(1,-1)$, so
only the plateau of $R_s= h/2e^2$ emerges. But for a large $M$ or a
small magnetic field $\phi$, $(\nu_{\uparrow},\nu_{\downarrow})$ may
be $(1,-3)$, $(3,-1)$, $(1,-5)$, $(5,-1)$, etc, then the plateaus of
$R_s= h/4e^2$, $h/6e^2$, etc, are also possible.

\begin{figure}
\includegraphics[width=8.5cm,totalheight=4cm]{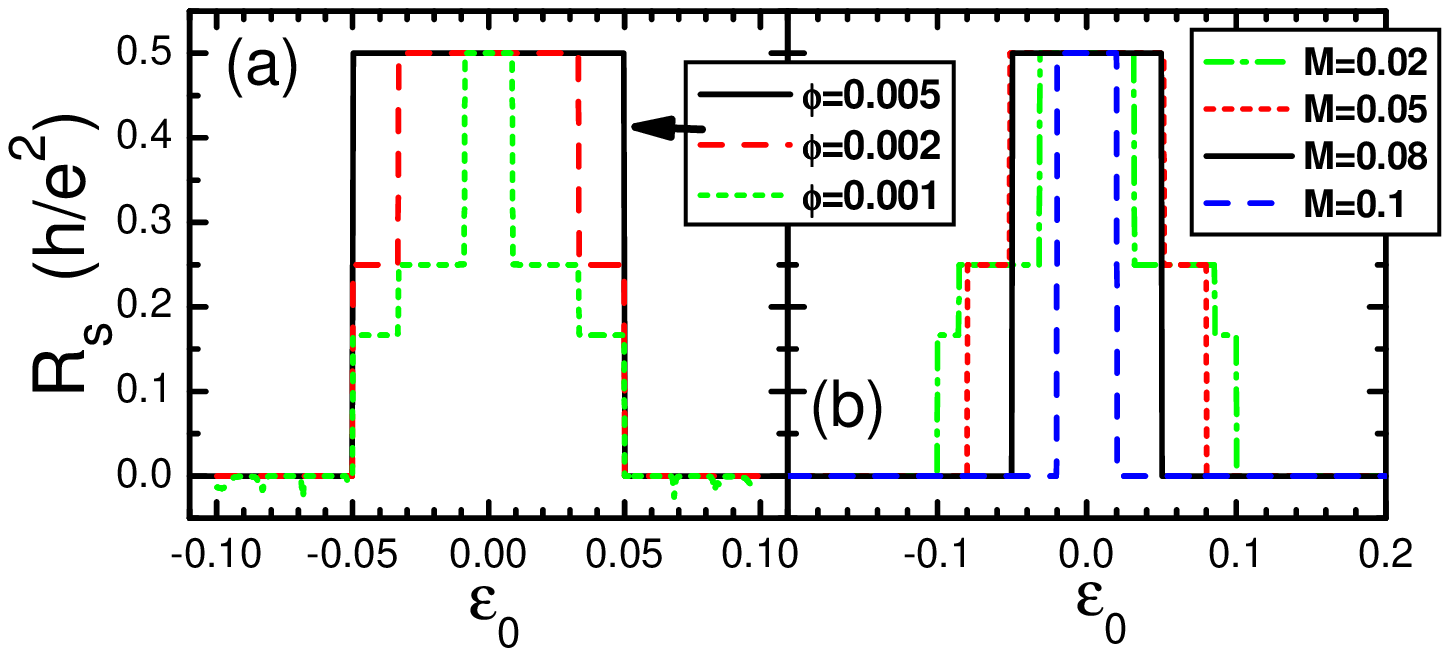}
\caption{ (Color online) (a) shows $R_s$ vs $\epsilon_0$ for
$M=0.05$ and (b) shows $R_s$ vs $\epsilon_0$ for $\phi=0.005$. The
parameter $N=80$.} \label{fig:4}
\end{figure}

\begin{figure}
\includegraphics[width=8.5cm,totalheight=4cm]{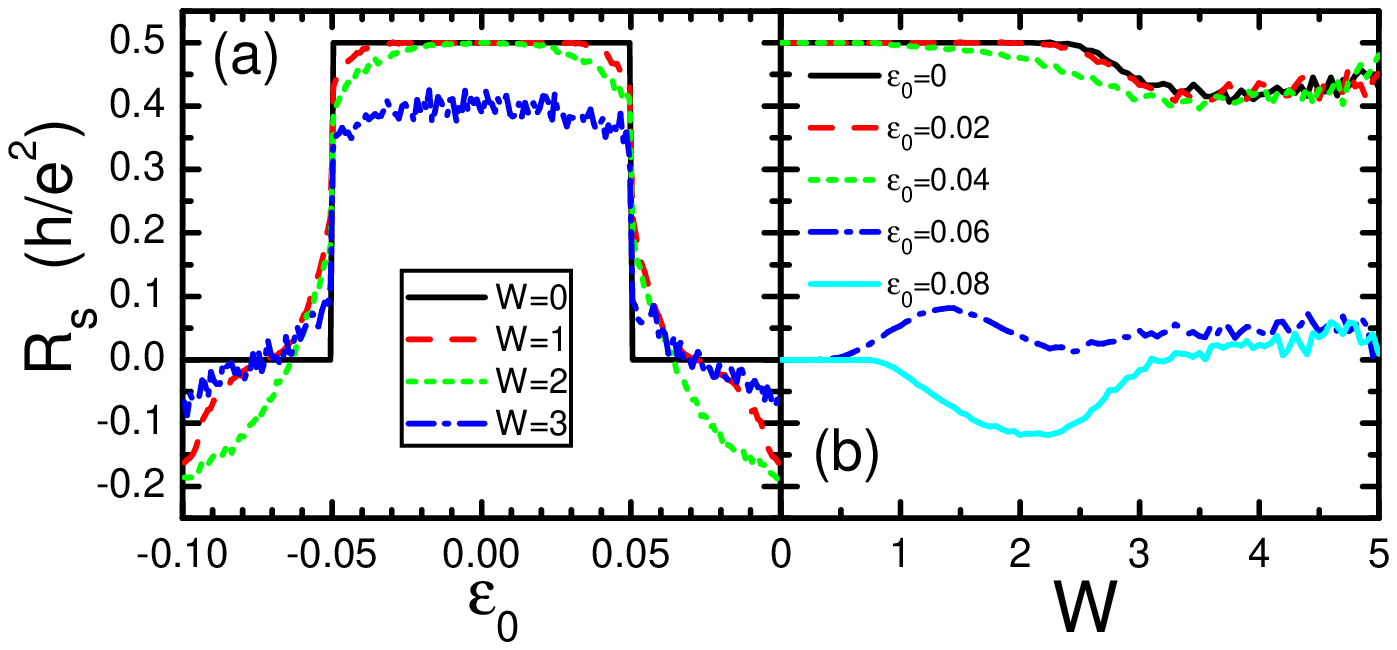}
\caption{ (Color online) $R_s$ vs $\epsilon_0$ (a) and $R_s$ vs the
disorder strength $W$ (b) with the parameters $N=40$, $\phi=0.007$,
and $M=0.05$. The curves in (a) and (b) are averaged over up to 1000
and 8000 random configurations, respectively. } \label{fig:5}
\end{figure}

Finally, we examine the disorder effect on the spin Hall resistance
$R_s$. Here we assume that the disorder only exists in the central
region (see dotted box in top right inset of Fig.1a). Due to the
disorder, the on-site energy $\epsilon_0 -\sigma M$ for each site
$i$ in the central region is changed to $\epsilon_0 +w_i -\sigma M$,
where $w_i$ is uniformly distributed in the range $[-W/2, W/2]$ with
the disorder strength $W$. Fig.5a shows $R_s$ versus the energy
$\epsilon_0$ at the different disorder strengths $W$ and Fig.5b
shows $R_s$ versus the disorder strength $W$ at different energies
$\epsilon_0$. The results show that the quantum plateaus of $R_s$
are very robust against the disorder because of the topological
invariance of the system. The quantum plateau maintains its
quantized value very well even when $W$ reaches 2 (see Fig.5a and
5b). Since the plateau is so robust and stable, its value can be
used as the standard for the spin Hall resistance. In addition, even
for a very large disorder strength $W$ (e.g. $W=5$ or larger), the
plateau value only slightly decreases while maintaining the plateau
structure (see Fig.5b). This is because although the disorder
strongly weakens the spin bias $\mu_{2\uparrow}-\mu_{2\downarrow}$,
it also weakens the longitudinal charge current $I_{13}$, so the
value of $R_s$ is affected less. This means that in the large
disorder limits ($W\rightarrow \infty$), although the QSHE is
broken, the SHE still holds.

In summary, we predict a new QSHE in the ferromagnetic graphene
film. Unlike the QSHEs studied so far, the origin of this QSHE is
not caused by the spin-orbit interaction. The results also show that
the system can exhibit the QSHE, the QHE, and the coexistence of the
QSHE and QHE, depending on the filling factors of the spin-up and
spin-down carriers. Due to the QSHE and QHE, both the longitudinal
and Hall resistances exhibit the plateau structures. The plateau
values (in the unit of $h/e^2$) are at $1/2$, $1/6$, $3/28$, ...,
for the longitudinal resistance and at $\pm 1/2$, $\pm 1/4$, $\pm
1/6$, $\pm 2/7$, ..., for the Hall resistance. In addition, the spin
Hall resistance has also investigated and found to be robust against
the disorder.

{\bf Acknowledgments:} This work was financially supported by
NSF-China under Grants Nos. 10525418, 10734110, and 10821403,
China-973 program and US-DOE under Grants No. DE-FG02- 04ER46124.
Q.F.S. gratefully acknowledges Prof. R. B. Tao for many helpful
discussions.

\end{document}